\documentclass[pra,aps,twocolumn,showpacs]{revtex4-1}
\usepackage{amsfonts}
\usepackage{amsmath}
\usepackage{amssymb}
\usepackage[dvips]{graphicx}

\begin{document}

\title{Dynamics of Entanglement Transfer Through Multipartite Dissipative Systems}
\date{\today}
\author{C.E. L\'opez$^{1,2}$, G. Romero$^{1,3}$ and J.C. Retamal$^{1,2}$}
\affiliation{$^{1}$Departamento de F\'{\i}sica, Universidad de Santiago de Chile, USACH, Casilla 307 Correo 2 Santiago, Chile
\\
$^{2}$Center for the Development of Nanoscience and Nanotechnology, 9170124, Estaci\'on Central, Santiago, Chile
\\
$^{3}$Departamento de Qu\'imica F\'isica, Universidad del Pa\'is Vasco - Euskal Herriko Unibertsitatea, Apdo. 644, 48080 Bilbao, Spain}
\pacs{03.65.Yz,
03.65.Ud, 03.67.Mn}

\begin{abstract}
We study the dynamics of entanglement transfer in a system composed of two initially correlated three-level atoms, each located in a cavity interacting with its own reservoir. Instead of tracing out reservoir modes to describe the dynamics using the master equation approach, we consider explicitly the dynamics of the reservoirs. In this situation, we show that the entanglement is completely transferred from atoms to reservoirs. Although the cavities mediate this entanglement transfer, we show that under certain conditions, no entanglement is found in cavities throughout the dynamics. Considering the entanglement dynamics of interacting and non-interacting bipartite subsystems, we found time windows where the entanglement can only flow through interacting subsystems, depending on
the system parameters.
\end{abstract}

\maketitle

\section{Introduction}
Entanglement has emerged as a central physical resource
for Quantum Information Theory~\cite{bouwmeester08}. For processing
quantum information, physical architectures are expected to be composed of multipartite quantum systems. Understanding how quantum entanglement is transferred between the parties has
motivated several contributions in recent years
~\cite{Cubitt05,Marcelo06-1,Eberly,Ficek08,Lopez08}. One issue is entanglement flow through individual parties and
the whole multipartite system~\cite{Cubitt05}. Another interesting	
problem is entanglement transfer between qubits and its relation
to energy~\cite{Marcelo06-1}. In addition, we have the study of the
entanglement transfer between non-interacting qubits, leading to
conservation rules for entanglement depending on how qubits are
initially correlated~\cite{Ficek08}. Entanglement transfer from
atoms to cavity modes leading to entanglement revivals has been
studied in \cite{Eberly}. Entanglement transference from two
cavities to their corresponding reservoirs allowed the description
of Entanglement Sudden Death as opposed to Entanglement Sudden Birth
\cite{Lopez08}.

An interesting problem in this context, is entanglement flow between parties in a multipartite system, including their dissipative mechanisms. In this work, we study the entanglement dynamics of two initially correlated atoms placed in two non-interacting leaky cavities, each connected to its own reservoir. To achieve this, we developed a hybrid analytical approach for finding the quantum dynamics of the atom-cavity-reservoir system. Unlike the master equation approach, our method allows us to include the reservoir dynamics, thus preventing information loss due to trace operations. We study the evolution of entanglement in different non-interacting bipartite subsystems, such as atom-atom, cavity-cavity and reservoir-reservoir. We show that the entanglement initially contained in the atomic subsystem is completely transferred into the reservoir-reservoir subsystem. Although cavities are the bridge connecting atoms to reservoirs, we show that they may not be entangled throughout the dynamics. Moreover, depending on the initial state, quantum dynamics may led to a situation where no entanglement exists in any of these three subsystems. In this case, we extend the study to other interacting and non-interacting subsystems.

This paper is organized as follows: in Section II we develop a hybrid analytical method to find the quantum dynamics. In Sec III, we study the dynamics of entanglement transfer. In Section I V we present our concluding remarks.

%%%%%%%%%%%%%%%%%%%%
\section{Atom-cavity-reservoir dynamics}
%%%%%%%%%%%%%%%%%%%%

Our model considers two independent subsystems, each formed by a three-level atom inside a leaky QED cavity. Each atom interacts with a single mode of frequency $\omega$ of the quantized electromagnetic field and a classical field with frequency $\nu$ in a Raman configuration, as shown in Fig.~\ref{scheme}. The quantum mode couples level $|g\rangle$ and $|c\rangle$ while classical field couples levels $|e\rangle$ and $|c\rangle$. Assuming no direct coupling between cavities, the dynamics of each  atom-cavity-reservoir system can be studied individually. Neglecting effects of spontaneous emission from leves $|c\rangle$ and $|e\rangle$, the Hamiltonian describing this system can be conveniently written, in the rotating wave approximation, as:

\begin{figure}[t]
\includegraphics[width=60mm]{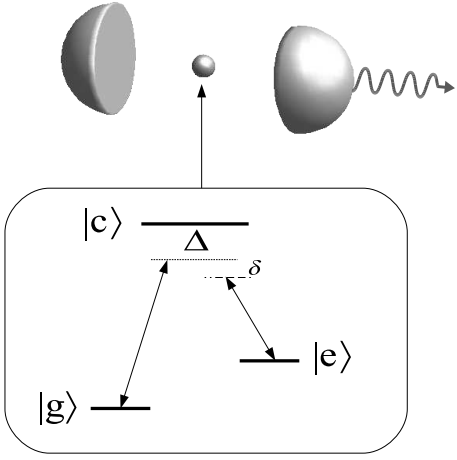}
\caption{Scheme of a three-level atom inside a cavity coupled to a reservoir.}
\label{scheme}
\end{figure}

\begin{eqnarray}
\hat{H}&=&\hbar \Delta |c\rangle \langle c|-\hbar \delta |e\rangle \langle e|-\hbar\sum_{k=1}^{N}(\omega-\omega_{k}) \hat{b}_k^{\dag }\hat{b}_k \notag \\
&&+\hbar \Omega (|c\rangle \langle e|+|e\rangle \langle c|)+ \hbar g \left( \hat{a} |c\rangle \langle g|+\hat{a}^{\dag } |g\rangle \langle c|\right) \notag \\
&&+\hbar\sum_{k=1}^{N}( g_{k}\hat{a}\hat{b}_{k}^{\dag}+g_{k}^{*}\hat{b}_{k}\hat{a}^{\dag }). \label{Hamiltonian2}
\end{eqnarray}

Here, $\hat{a}$$(\hat{a}^{\dag })$ annihilates (creates) a photon with frequency $\omega$ in the cavity mode, operator $\hat{b}_{k}$$(\hat{b}^{\dag }_{k})$
annihilates (creates) a photon with frequency $\omega_k$ in $k$-th mode  of the reservoir. We have defined $\Delta=\omega_{cg}-\omega$, with $\omega_{cg}$ is the frequency difference between levels $|c\rangle$ and $|g\rangle$, detuning $\delta=\omega_{ce}-\nu-\Delta$. The coupling strength between the classical field and the atom is $\Omega$  and we have explicitly written the system-bath interaction with a linear coupling with coupling constant $g_k$.  In the usual master equation approach, considering the Markov approximation and an infinite number of bath oscillators, we can describe the dynamics of the atom-cavity system by
\begin{equation}
\dot{\hat{\rho}}=-i[\hat{H'},\hat{\rho}]+\frac{\kappa}{2}\left( 2 \hat{a}^{\dag}\hat{\rho}\hat{a}-\hat{a}^{\dag}\hat{a}\hat{\rho}-\hat{\rho}\hat{a}^{\dag}\hat{a}\right),
\label{mastereq}		
\end{equation}
where $\hat{H'}$ corresponds to Hamiltonian~(\ref{Hamiltonian2}) without terms involving the reservoir.

In such an approach, only atom-cavity dynamics is described, while the reservoirs are traced out. This results in loss of quantum correlations between
reservoirs and other parties.

In what follows, we will develop a method to find the dynamics, including explicitly the reservoir's degrees of freedom. In this manner we preserve all the information about quantum correlations in the system

To address this problem, we develop a hybrid analytical approach to find  the dynamics. First, we will follow the method described for obtaining the quantum dynamics and the entanglement properties of inhomogeneously-coupled
systems~\cite{Lopez07-1,Lopez07-2}. This method consists of inspecting the Hilbert space occupied for the quantum system throughout the evolution and implementing a truncation criteria based on a probabilistic argument. Then, to find explicitly analytical expressions for the quantum dynamics, we follow the well-known
quantum trajectory method~\cite{DiFidio08}. 

Let us consider the case of a single initial excitation contained in  the atomic subsystem. That is,
\begin{equation}
|\psi_{0}\rangle= |e\rangle_{a}\otimes |0\rangle_{c}\otimes |\mathbf{\mathbf{\bar{0}}}\rangle_{r} \equiv |e0\mathbf{\bar{0}}\rangle,
\label{initial}
\end{equation}
where, $|0\rangle_{c}$ corresponds to the vacuum of the cavity and
$|\mathbf{\bar{0}}\rangle_{r}\equiv \prod_{k} |0_{k}\rangle $
denotes the vacuum of the reservoir. Now we have to look for  the
accessible Hilbert space for the compound system when starting from
this initial condition. We can find this by following the action of
Hamiltonian $\hat{H}_{II}$ on the initial state $|\psi_{0}\rangle$.
It is not difficult to realize that a portion of the Hilbert space
connected by Hamiltonian (\ref{Hamiltonian2}) is given by states
$\{|e0\mathbf{\bar{0}}\rangle, |c0\mathbf{\bar{0}}\rangle,
|g1\mathbf{\bar{0}}\rangle \}$. The one photon excitation in state
 $|g1\mathbf{\bar{0}}\rangle$ is transferred to bath modes through
 Hamiltonian (\ref{Hamiltonian2}) as follows:
\begin{equation}
\hat{H}_{II} |g1\mathbf{\bar{0}}\rangle=\hbar g  |c0\mathbf{\bar{0}}\rangle + \hbar N_{0}  |g0\mathbf{\bar{1}}_0 \rangle,
\label{act3}
\end{equation}
where we have defined the state:
\begin{equation}
|\mathbf{\bar{1}}_{0}\rangle_{r} \equiv \frac{1}{N_{0}} \sum_{k=1}^{n} g_{k} |1_{k}\rangle_{r}.  \label{onek}
\end{equation}
with $N_{0}=\sqrt{\sum_{k=1}^{n} |g_{k}|^2}$, and
$|1_{k}\rangle_{r}$ corresponds to the state having one photon in
the $k$-th reservoir mode and zero photon in the remaining modes. The state in
Eq.~(\ref{onek}) is a collective state of the reservoir having a
single excitation. The $|g0\bar{1}_0 \rangle$ collective state
evolves under the interaction part in
Hamiltonian~(\ref{Hamiltonian2}), back to states
$|g1\mathbf{\bar{0}}\rangle$ and $|c0\mathbf{\bar{0}}\rangle$.
However, this is no longer true when considering the action of the
reservoir Hamiltonian on reservoir states. Taking the vacuum
state of the reservoir, we have
\begin{eqnarray}
\sum_{k}^{N}(\omega-\omega_{k})\hat{b}_{k}^{\dag}\hat{b}_{k}|\mathbf{\bar{0}}\rangle_{r} & = & 0.
\end{eqnarray}
For $|\bar{1}_{0}\rangle_{r}$, free energy term
leads to
\begin{eqnarray}
\mid \Phi_1 \rangle=
\frac{1}{N_{0}}\sum_{k}^{N}g_{k}(\omega-\omega_{k})|1_{k}\rangle_{r}.
\label{1}
\end{eqnarray}
Although this state is different from $|\bar{1}_{0}\rangle_{r}$ in Eq.(\ref{onek}), it can be written as a superposition of this state and a state $|\bar{1}_{1}\rangle_{r}$ orthogonal to $|\bar{1}_{0}\rangle_{r}$. That is,
\begin{equation}
|\mathbf{\bar{1}}_{1}\rangle_{r}=\frac{1}{N_{1}}\big[|\Phi_{1}\rangle- \langle\mathbf{\bar{1}}_{0}|\Phi_{1}\rangle|\mathbf{\bar{1}}_{0}\rangle_{r}\big],
\label{}
\end{equation}
where $N_{1}=(\langle \Phi_{1}|\Phi_{1}\rangle-|\langle
\mathbf{\bar{1}}_{0}|\Phi_{1}\rangle|^2)^{1/2}$.

The new generated state $|\mathbf{\bar{1}}_{1}\rangle_{r}$ leads to other
orthogonal states having one excitation through the action of the
reservoir terms of the Hamiltonian~\cite{Lopez07-1}. In consequence,
the accessible Hilbert space for the overall system initially
prepared in state~({\ref{initial}}), can be written in a
collective basis spanned by the set of orthogonal states
\begin{eqnarray}
&\{&|e0\mathbf{\bar{0}}\rangle,|c0\mathbf{\bar{0}}\rangle,|g1\mathbf{\bar{0}}\rangle,|g0\mathbf{\bar{1}}_{0}\rangle,|g0\mathbf{\bar{1}}_{1}\rangle,|g0\mathbf{\bar{1}}_{2}\rangle,\dots\}.
\label{basis}
\end{eqnarray}
Therefore, the atom-cavity-reservoir system governed by
Hamiltonian~(\ref{Hamiltonian2}) evolves to
\begin{eqnarray}
|\psi_{t}\rangle&=&E_{t}|e0\mathbf{\bar{0}}\rangle+C_{t}|c0\mathbf{\bar{0}}\rangle+G_{t}|g1\mathbf{\bar{0}}\rangle \notag\\
&&+R_{0,t}|g0\mathbf{\bar{1}}_{0}\rangle+R_{1,t}|g0\mathbf{\bar{1}}_{1}\rangle+ \dots,
\label{psit}
\end{eqnarray}
where $R_{j,t}$ is the time-dependent probability amplitude of the
state $|g0\mathbf{\bar{1}}_{j}\rangle$. This state can conveniently written
defining  one-excitation collective state for the reservoir
\begin{equation}
|\mathbf{\bar{1}}\rangle=\frac{1}{R_{t}}\left(R_{0,t}|\mathbf{\bar{1}}_{0}\rangle+R_{1,t}|\mathbf{\bar{1}}_{1}\rangle+
\dots\right), \label{collective}
\end{equation}
where,
\begin{equation}
R_{t}=\sqrt{|R_{0,t}|^2+|R_{1,t}|^2+\dots},
\label{R}
\end{equation}
corresponds to the probability amplitude of having one excitation in
the reservoir modes.

In this manner, the dynamics of the atom-cavity-reservoir system can
be described in terms of a three-qubit system in the basis:
$\{|e\rangle,|g\rangle\} \otimes \{|0\rangle,|1\rangle\} \otimes
\{|\mathbf{\mathbf{\bar{0}}}\rangle,|\mathbf{\bar{1}}\rangle \}$.
Notice that we have assumed the high detuning regime $(\Delta \gg
\Omega,g)$, so the electronic level $|c\rangle_{a}$ is only
virtually populated and the evolution described in Eq.~(\ref{psit})
can now be written as
\begin{equation}
|\psi_{t}\rangle=E_{t}|e0\mathbf{\bar{0}}\rangle+G_{t}|g1\mathbf{\bar{0}}\rangle+R_{t}|g0\mathbf{\bar{1}}\rangle.
\label{psit2}
\end{equation}
Probability amplitudes can be obtained by numerical
diagonalization of Hamiltonian~(\ref{Hamiltonian2}) where it is found that $C_{t}\sim 0$. It is worth to notice that this form of finding the quantum dynamics has close relation with the Weisskopf-Wigner procedure (see for example ref.~\cite{Scullybook}) where the reservoir is considered throughout the dynamics.

We are interested in finding analytical expressions for the temporal
dependent coefficients in the wave function in Eq. (\ref{psit2}).
Unfortunately, this is not possible by solving the Schr\"odinger
equation. However, if we know the reduced dynamics for the atom-cavity
system, we can get information about the probability amplitude
$R_t$. The reduced atom-cavity dynamics could be obtained by solving
the corresponding master equation, or alternatively by using the
quantum trajectory method (see for example~\cite{DiFidio08} and
references within). By using this approach, we obtain
the reduced density matrix for the atom-cavity system as:
\begin{eqnarray}
\hat{\rho}(t)&=&|E_{t}|^2|e0\rangle \langle e0|+|G_{t}|^2|g1\rangle \langle g1|\notag \\
&&+E_{t}G^{*}_{t}|e0\rangle \langle g1|+E^{*}_{t}G_{t}|g1\rangle \langle e0|\notag \\
&&+|R_{t}|^2 |g0\rangle \langle g0| \label{rho1}
\end{eqnarray}
The probability amplitudes obtained through the quantum trajectory approach are
\begin{eqnarray}
E_{t}&=&\{\cos{(\bar{\Omega} t)}+\frac{\kappa}{4\bar{\Omega}}\sin{(\bar{\Omega}t)}\}
e^{-\frac{1}{4}\kappa t}, \notag \\
G_{t}&=&\frac{i g_{\rm eff}}{ \bar{\Omega}}\sin{(\bar{\Omega}t)}e^{-\frac{1}{4}\kappa
t}, \label{coeft}
\end{eqnarray}
where $4\bar{\Omega}^2=4g_{\rm eff}^2-\kappa^2/4$ and $g_{\rm eff}=g
\Omega / \Delta$. For simplicity, here we have set $\delta=(g^2-\Omega^2)/\Delta$. From these expressions we obtain the probability
amplitude $R_t$ as:
\begin{eqnarray}\label{Rt}
|R_{t}|&=&\sqrt{1-|E_{t}|^2-|G_{t}|^2}.
\end{eqnarray}

On the other hand, since $R_{t}$ is by definition a real-positive number ($|R_{t}|= R_{t}$), we can say that
the dynamics is given by Eq.(\ref{psit2}) with $E_{t}$,$G_{t}$ and $R_{t}$, defined in Eqs.~(\ref{coeft}) and (\ref{Rt}).

%%%%%%%%%%%%%%%%%%%%%%
\section{Dynamics of Entanglement transfer}
%%%%%%%%%%%%%%%%%%%%%%

\begin{figure}[t]
\includegraphics[width=60mm]{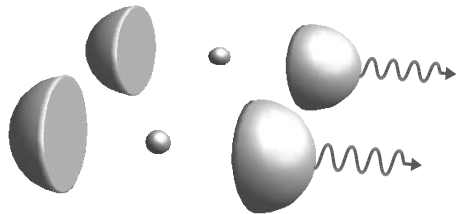}
\caption{Scheme of two initially entangled atoms, each located in a leaky cavity.}
\label{scheme2}
\end{figure}

Having described the dynamics of the atom-cavity-reservoir subsystem, we now focus on the problem of two initially entangled atoms, each located in a leaky cavity, as shown in Fig.~\ref{scheme2}. This entangled state between distant atoms have received much attention in the last years due to its possible applications in Quantum Communication~\cite{Cirac97,Serafini06,Yu07,Yu08,Morigi08-2,Wang09,Vogel09}. In this sense, the present study might be relevant. On the other hand, this coherent manipulation of atoms and cavities could also be extended into a trapped ions system. For example in the experimental setup described in reference~\cite{Russo09}, a single trapped ion is coupled to a high-finesse cavity. Further improvements in this experiment could consider a second cavity coupled to a second ion. In this case, the same dynamics we are describing in this section can be found in this system. Entanglement between the electronic levels of the ions can be prepared using the center-of-mass mode.
 
In our system, since there is no interaction between both atom-cavity-reservoir subsystems, the only link is provided by the entanglement between the atoms.
Within this scenario, we are interested in the study of how the
entanglement, initially shared by the two atoms, is transferred to
other parties. To do this, we first consider an initial entangled
atomic state given by
\begin{equation}
|\Psi_{0}\rangle=\left(\alpha|gg\rangle_{12}+\beta |ee\rangle_{12} \right)|00\rangle_{12}|\mathbf{\bar{0}}\mathbf{\bar{0}}\rangle_{12},
\label{initial2}
\end{equation}
where cavities and reservoirs are in the vacuum state. Following the
previous single atom-cavity-reservoir analysis, the initial state
$|\Psi_{0}\rangle$ evolves to:
\begin{equation}
|\Psi_{t}\rangle=\alpha|g0\mathbf{\bar{0}}\rangle_{1}\otimes|g0\mathbf{\bar{0}}\rangle_{2}+\beta|\psi_{t}\rangle_{1}\otimes|\psi_{t}\rangle_{2},
\label{evoltpsi}
\end{equation}
where $|\psi_{t}\rangle$ is given in Eq.~(\ref{psit2}). By tracing
out the degrees of freedom of cavities and reservoirs we are led
to the atom-atom dynamics:
\begin{eqnarray}
\rho_{a_{1}a_{2}}&=&\beta^2|E_{t}|^4|ee\rangle \langle ee|+\alpha\beta|E_{t}|^2 \left(|ee\rangle \langle gg|+|gg\rangle \langle ee| \right) \notag \\
&&+\beta^2|E_{t}|^2\left(|G_{t}|^2+|R_{t}|^2\right) \left(|eg\rangle \langle eg|+|ge\rangle \langle ge| \right) \notag \\
&&+(\alpha^2+\beta^2 \left(|G_{t}|^2+|R_{t}|^2\right)^2 )|gg\rangle \langle gg|.
\label{atomatommatrix}
\end{eqnarray}
Atomic entanglement can be evaluated by using the
concurrence~\cite{Wootters98}, leading to:
\begin{equation}
\mathcal{C}_{a_{1}a_{2}}(t)=\max\{0,-2\lambda^{a_{1} a_{2}}_{-}\},
\label{conca1a2}
\end{equation}
where the negative eigenvalue of the partial transpose matrix \cite{Peres,Horodecki} of
$\rho_{a_{1}a_{2}}$ is given by:
\begin{equation}
\lambda^{a_{1} a_{2}}_{-}=\beta |E_{t}|^2 \left(\beta \left(1-|E_{t}|^2 \right)-\alpha \right).
\label{lambdaa1a2}
\end{equation}
The entanglement flow from atoms to other parties is obtained from
cavity-cavity and reservoir-reservoir reduced systems, respectively.
The reduced cavity-cavity system is described by:
\begin{eqnarray}
\rho_{c_{1}c_{2}}&=&\beta^2|G_{t}|^4|11\rangle \langle 11|+\alpha\beta|G_{t}|^2 \left(|00\rangle \langle 11|+|11\rangle \langle 00| \right) \notag \\
&&+\beta^2|G_{t}|^2\left(|E_{t}|^2+|R_{t}|^2\right) \left(|10\rangle \langle 10|+|01\rangle \langle 01| \right) \notag \\
&&+(\alpha^2+\beta^2 \left(|E_{t}|^2+|R_{t}|^2\right)^2 )|00\rangle \langle 00|,
\label{cavcavmatrix}
\end{eqnarray}
while for the reduced reservoir-reservoir system we have:
\begin{eqnarray}
\rho_{r_{1}r_{2}}&=&\beta^2|R_{t}|^4 |\mathbf{\bar{1}}\mathbf{\bar{1}} \rangle \langle \mathbf{\bar{1}}\mathbf{\bar{1}}|+\alpha\beta|R_{t}|^2 \left(|\mathbf{\bar{0}} \mathbf{\bar{0}} \rangle \langle \mathbf{\bar{1}}\mathbf{\bar{1}}|+|\mathbf{\bar{1}} \mathbf{\bar{1}} \rangle \langle \mathbf{\bar{0}} \mathbf{\bar{0}} | \right) \notag \\
&&+\beta^2|R_{t}|^2 \left(|E_{t}|^2+|G_{t}|^2\right) \left(|\mathbf{\bar{1}} \mathbf{\bar{0}}\rangle \langle \mathbf{\bar{1}} \mathbf{\bar{0}}|+|\mathbf{\bar{0}} \mathbf{\bar{1}} \rangle \langle \mathbf{\bar{0}} \mathbf{\bar{1}}| \right) \notag \\
&&+(\alpha^2+\beta^2 \left(|E_{t}|^2+|G_{t}|^2\right)^2 )|\mathbf{\bar{0}} \mathbf{\bar{0}}\rangle \langle \mathbf{\bar{0}} \mathbf{\bar{0}}|.
\label{resresmatrix}
\end{eqnarray}
The corresponding entanglement in these subsystems can also be calculated
using the concurrence. The concurrences for cavities and reservoirs, respectively, are given by:
\begin{eqnarray}
\mathcal{C}_{c_{1}c_{2}}(t) &=& \max\{0,-2\lambda^{c_{1} c_{2}}_{-}\} , \label{concc1c2}\\
\mathcal{C}_{r_{1}r_{2}}(t) &=& \max\{0,-2\lambda^{r_{1} r_{2}}_{-}\}
\label{concr1r2}
\end{eqnarray}
where the negatives eigenvalues of partial transposed matrices are
given by:
\begin{eqnarray}
\lambda^{c_{1} c_{2}}_{-} &=& \beta |G_{t}|^2 \left(\beta \left(1-|G_{t}|^2 \right)-\alpha \right)\label{lambdac1c2} \\
\lambda^{r_{1} r_{2}}_{-} &=& \beta |R_{t}|^2 \left(\beta \left(1-|R_{t}|^2 \right)-\alpha \right). \label{lambdar1r2}
\end{eqnarray}

\begin{figure}[t]
\includegraphics[width=80mm]{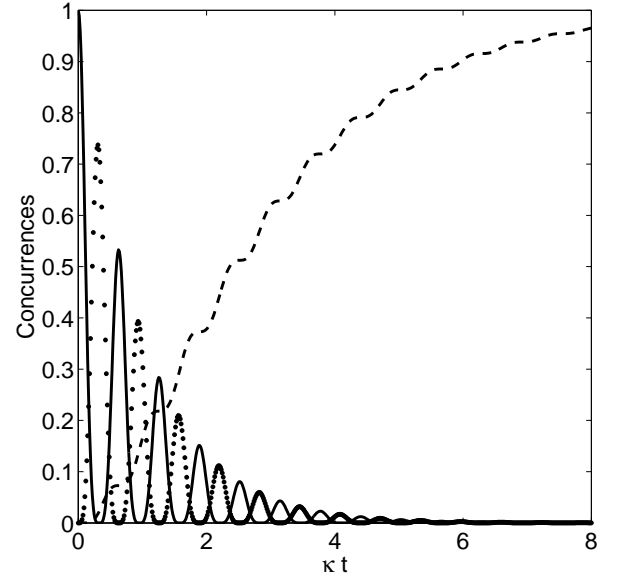}
\caption{Evolution of concurrence of subsystems: $a_1\otimes a_2$ (solid line); $c_1\otimes c_2$ (dashed line) and $r_1\otimes r_2$ (dot-dashed line) for an initial state of the form of Eq.~(\ref{initial2}) with $\alpha=\beta=1/\sqrt 2$ and $g_{\rm{eff}}=5\kappa$.}
\label{Dynamics1}
\end{figure}

We are now in a position to investigate the evolution of entanglement
for different subsystems. Disentanglement dynamics depends on probability amplitudes $\alpha$ and $\beta$, as well as the
decay constant $\kappa$ compared to the effective coupling
$g_{\rm eff}$. For example, an important case is an initial
maximally entangled state. Fig.~\ref{Dynamics1}  shows the
entanglement evolution for atoms, cavities and reservoir subsystems,
starting from this state with $\alpha =\beta$ for $g_{\rm
eff}=5\kappa$. For the chosen parameters, we observe that the entanglement exhibits asymptotic decay, with oscillations of atomic entanglement and cavities
entanglement to finally be completely transferred into the
reservoirs. These oscillations are expected because for $g_{\rm
eff}>\kappa$, there are many energy exchange between atomic and
cavity subsystems	 before the energy is completely transferred to
reservoirs. In addition, the entanglement between reservoirs appears
at the same time as entanglement between atoms begins to decrease. This
regime is reminiscent from what happens for an initial maximally
entangled state under the action of a dissipative environment.

For cavities with no atoms and the initial state given by
$|\Psi\rangle=(c_0|00\rangle+c_1|11\rangle)|\mathbf{\bar{0}}
\mathbf{\bar{0}}\rangle$, it is well known that entanglement
disappears at a finite time if $c_1>c_0$~\cite{Marcelo06}.
Otherwise, if $c_0 > c_1$, the entanglement decays asymptotically.
Moreover, in ref.~\cite{Lopez08}, finite-time disentanglement, known
as, {\it entanglement sudden death} (ESD)~\cite{Zyczkowski01,Diosi03,Yu04} is necessarily linked to
a sudden birth of entanglement between reservoirs, called {\it
entanglement sudden birth} (ESB). This behavior can also occur
in our system by considering unbalanced states with $\beta >
\alpha$.
\begin{figure}[t]
\includegraphics[width=85mm]{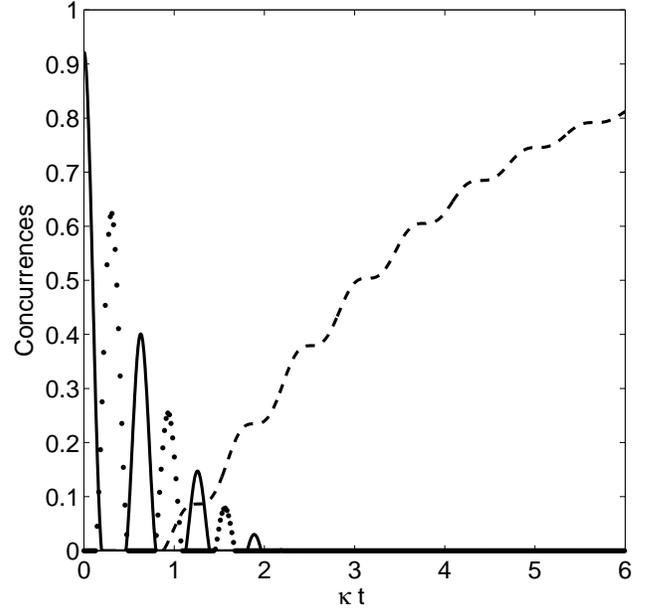}
\caption{Evolution of concurrence of subsystems:$a_1\otimes a_2$(solid line); $c_1\otimes c_2$ (dots) and $r_1\otimes r_2$ (dashed line) for an initial state of the form of Eq.~(\ref{initial2}) with $\beta=1.5 \alpha $ and $g_{\rm{eff}}=5\kappa$.}
\label{strongc}
\end{figure}
In Fig.~\ref{strongc}, the evolution of the entanglement contained in
the three subsystems is shown for $\beta > \alpha$ and $g_{\rm
eff}=5\kappa$. We realize that entanglement still experiences oscillations between cavities and atoms, however, entanglement experiences several sudden deaths and entanglement sudden revivals (ESR), whereas the reservoirs experience a sudden birth of entanglement.

These behaviors can in principle be quantitatively understood from
Eqs.~(\ref{lambdaa1a2}), ~(\ref{lambdac1c2}) and
~(\ref{lambdar1r2}). For arbitrary effective coupling constant
$g_{\rm eff}$ and amplitudes $\alpha$ and $\beta$, it is not
possible to calculate the times at which ESD, ESR or ESB appears.
But for the case of reservoir entanglement Eq.~(\ref{lambdar1r2})
says that for $\alpha > \beta$ there will always be an entanglement
birth at time $t=0$. However, for $\beta > \alpha $ the reservoirs
will remain unentangled unless:
\begin{eqnarray}
|R_{t}|^2 > 1-\frac{\alpha}{\beta}.\label{condr1r2}
\end{eqnarray}
We observe this is the behavior in Fig ~\ref{strongc} where
ESB appears for a $t_{\rm ESB}>0$.

Some analytical calculations can be carried out in
special regimes of $g_{\rm eff}$ as compared to	 $\kappa$. In the strong coupling regime, when $g_{\rm eff} \gg \kappa$ we have
$\bar{\Omega}\approx g_{\rm eff}$, so that:
\begin{eqnarray}
|E_{t}|^2 &\approx& \cos^{2}{(g_{\rm eff} t)}e^{-\frac{1}{2}\kappa t}  \\
|G_{t}|^2 &\approx& \sin^{2}{(g_{\rm eff} t)}e^{-\frac{1}{2}\kappa t}\\
|R_{t}|^2 &\approx&1-e^{-\frac{1}{2}\kappa t} \label{}
\end{eqnarray}
From Eqs. ~(\ref{lambdaa1a2}) and ~(\ref{lambdac1c2}) we can explain
the entanglement death and revivals zones for atoms and cavities in
Fig. ~\ref{strongc}. Conditions for disentanglement in both
subsystems are given by:
\begin{eqnarray}
&1-\cos^{2}{(g_{\rm eff} t)}e^{-\frac{1}{2}\kappa t} \geq \frac{\alpha}{\beta} \label{cond1}\\
&1-\sin^{2}{(g_{\rm eff} t)}e^{-\frac{1}{2}\kappa t} \geq
\frac{\alpha}{\beta} \label{cond21}
\end{eqnarray}
\begin{figure}[t]
\hspace*{-1.0cm}
\includegraphics[width=95mm]{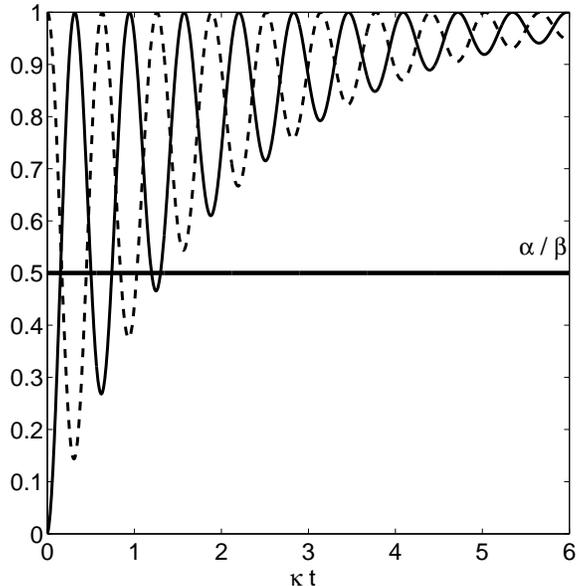}
\caption{Evolution of Eqs. (\ref{cond1})  (solid line) and (\ref{cond21}) (dashed line) for $g_{\rm{eff}}=5 \kappa$. Dot-dashed line shows the value of $\alpha/\beta$.}
\label{cond}
\end{figure}
In the strong coupling regime, we find time windows with no
entanglement between atoms, that is, time windows between ESD and
ESR times. These time windows for atoms happen at time intervals
that can be different from those for cavities, or can overlap, depending on the ratio $\alpha/\beta$. At the same time the entanglement in both subsystems extinguish definitely because of dissipation given by the attenuation factor in the previous
inequalities. This can be observed in Fig.{\ref{cond}, where
evolutions of Eqs.~(\ref{cond1}) and (\ref{cond21}) are shown
compared to $\alpha/\beta$. Here we see that the attenuation
factor $\exp{(-\kappa t /2)}$ makes it possible to satisfy conditions
(\ref{cond1}) and (\ref{cond21}), leading to complete disentanglement
of atoms and cavities. In addition, from Eq.
(\ref{condr1r2})  we can evaluate  the time for ESB in the reservoir subsystem, which
is given by:
\begin{equation}
t_{\rm{ESB}} \approx 2\frac{1}{\kappa} \ln{\bigg[\frac{\beta}{\alpha}\bigg]}.
\label{esbtime}
\end{equation}
This  time is twice the found for two entangled dissipative
cavities studied in ~\cite{Lopez08}. In the present case ESB also happens if $\beta
> \alpha$.
\begin{figure}[t]
\hspace*{-1.3cm}
\includegraphics[width=85mm]{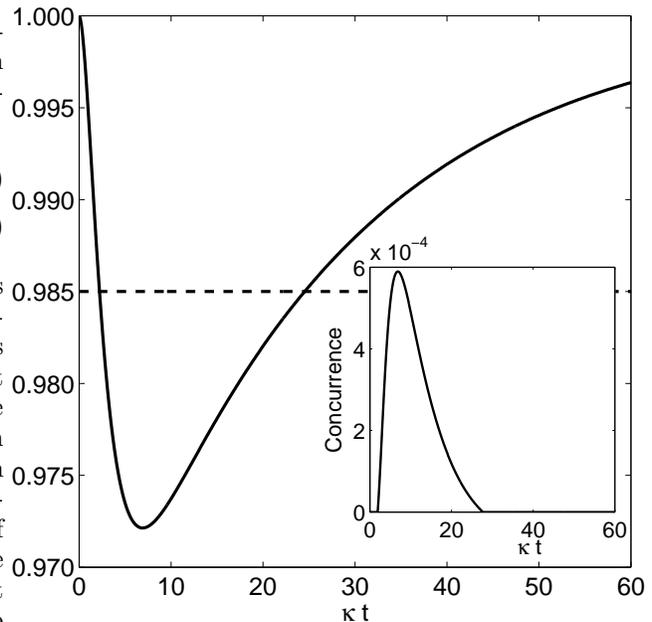}
\caption{Evolution of the left-hand side of Eq.~(\ref{condg})  (solid
line) for $g_{\rm eff}=0.1 \kappa$. Dashed line shows $\alpha/\beta=
0.985$. The inset shows the concurrence $\mathcal{C}_{c_{1}c_{2}}(t)
$ for the cavity-cavity subsystem.} \label{cond2}
\end{figure}

On the other hand, in the weak coupling regime $g_{\rm eff} \ll \kappa$, we find that:
\begin{eqnarray}
|E_{t}|^2 &\approx &(1+4 \gamma^2 ) e^{-4\gamma^2 \kappa t}-4\gamma^2 e^{-\kappa t+4\gamma^2 \kappa t} \label{et} \\
|G_{t}|^2 &\approx& 4\gamma^2 (e^{-4\gamma^2 \kappa t}+e^{-\kappa t+4\gamma^2 \kappa t}-2 e^{-\kappa t /2}) \label{gt} \\
|R_{t}|^2 &\approx &1-(1+8 \gamma^2 )e^{-4 \gamma^2 \kappa t}+8
\gamma^2 e^{-\kappa t /2},\label{rt}
\end{eqnarray}	
where we have considered only up to second order in $\gamma=g_{\rm
eff} / \kappa$. From these equations we observe that unlike the
strong coupling regime, in the weak coupling regime the entanglement
dynamics is not oscillatory, that is, no entanglement revivals can
be found in atoms or cavities. We expect that disentanglement
between atoms be followed by entanglement birth in reservoirs.
However entanglement between cavities depends on $\gamma$ and
the ratio $\alpha/\beta$. This can be understood by considering
Eqs.~(\ref{lambdac1c2}) and (\ref{gt}), such that the condition for
disentanglement in the cavities is given by:
\begin{equation}\label{condg}
1-4\gamma^2 (e^{-4\gamma^2 \kappa t}+e^{-\kappa t+4\gamma^2 \kappa t}-2 e^{-\kappa t /2}) \ge \frac{\alpha}{\beta}
\end{equation}
The evolution of the left-hand side of the inequality is shown in
Fig. \ref{cond2}. According to (\ref{condg}), this figure shows that
cavities are entangled depending on the ratio $\alpha/ \beta$. More
precisely, cavities get entangled only while the curve is below the
value of $\alpha/ \beta$. In particular, for the case of $\alpha/
\beta=0.985$ (dashed line) cavities get entangled only when the
curve is below the dashed line, as shown in the inset where the
cavity-cavity concurrence is plotted. When the ratio $\alpha/ \beta$
is set to a value below the minimum of the left hand side in
(\ref{condg}) cavities never become entangled. The particular
case of Fig \ref{cond2} happens for $\alpha/\beta < 0.972$. In
such a case the entanglement initially contained in the atomic
subsystem is transferred directly to the reservoirs without
entangling the cavities. This feature of entanglement dynamics can even be found outside the weak coupling regime as a function of  $\gamma$ and $\alpha / \beta$. We can distinguish an entangled phase and an unentangled phase throughout the dynamics, which can be obtained from Eqs. (\ref{concc1c2}) and (\ref{lambdac1c2}). Fig.~\ref{diagram} shows the two phases for the cavities entanglement dynamics as a function of $\gamma$ and $\alpha/\beta$.

In the case of atom and reservoir subsystems, the non oscillatory behavior is
shown in Fig~\ref{weakc} for $\gamma=0.1$ and $\alpha/\beta=2/3$. In
such a case, the atom subsystem exhibits ESD, while the reservoir
subsystem exhibits ESB. The time at which ESD and ESB happen depends
on the ratio $\alpha/\beta$ for a fixed $\gamma$. In the weak
coupling regime, and for values of $\gamma$ that allows to neglect
corrections in $\gamma^2$ in both Eqs. (\ref{et}) and (\ref{rt}), we have:
\begin{eqnarray}
t_{\rm{ESD}}&\approx&\left(\frac{1}{4 \gamma^2}\right)\frac{1}{\kappa} \ln{\bigg[\frac{\beta}{\beta-\alpha}\bigg]},\label{tweakD} \\
t_{\rm{ESB}}&\approx&\left(\frac{1}{4 \gamma^2}\right)\frac{1}{\kappa}\ln{\bigg[\frac{\beta}{\alpha}\bigg]}. \label{tweakB}
\end{eqnarray}
\begin{figure}[t]
\vspace*{-0.5cm}
\hspace*{-0.5 cm}
\includegraphics[width=95mm]{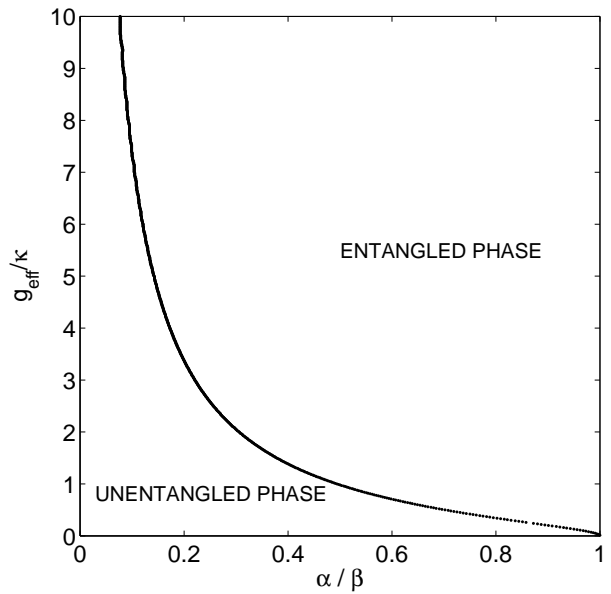}
\caption{Diagram showing the entangled/unentangled phases of the cavity-cavity subsystem as a function of coefficients $g_{\rm eff}/\kappa$ and $\alpha/\beta$.}
\label{diagram}
\end{figure}
\begin{figure}[t]
\hspace*{-1.3cm}
\includegraphics[width=87mm]{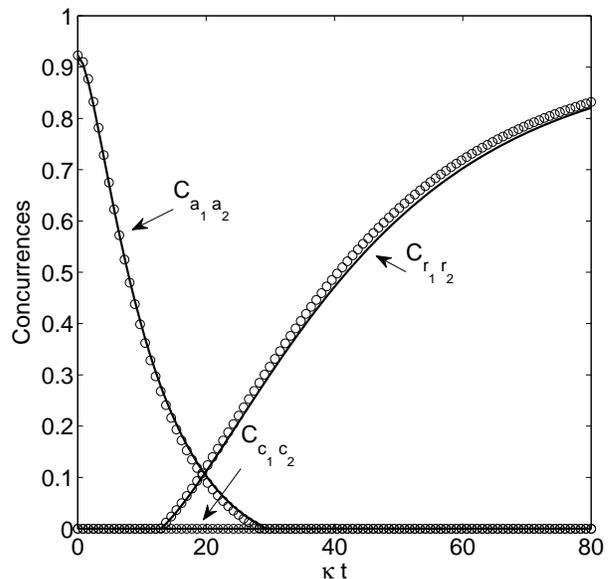}
\caption{Evolution of concurrence of subsystems $a_1\otimes a_2$,  $c_1\otimes c_2$ and $r_1\otimes r_2$, using Eqs.~(\ref{et})-(\ref{rt}) (solid lines) and exact calculations (circles). Parameter are $\alpha/\beta=2/3$ and $g_{\rm{eff}}=0.1
\kappa$.} \label{weakc}
\end{figure}
From  these Eqs. we observe that the ESB time can occur before, simultaneously or after the ESD. For example, in Fig.~\ref{tw}, entanglement dynamics is shown for $\gamma=0.1$ and $\beta=3\alpha$. In such a figure we observe that there is a time window for which no entanglement is found in the three subsystems. Using Eqs.(\ref{tweakD}) and (\ref{tweakB}), we can calculate the size of this time window, leading to:
\begin{equation}
\Delta t_W \approx \left(\frac{1}{4 \gamma^2}\right)\frac{1}{\kappa}\ln{\bigg[\frac{\beta}{\alpha}-1\bigg]}.
\end{equation}
We realize from this expression that the time window can exist only if $\beta > 2 \alpha$ and increases in size, as well as coupling strength $g_{\rm eff}$ decreases.  
\begin{figure}[t]
\vspace*{-0.5cm}
\hspace*{-0.5cm}
\includegraphics[width=87mm]{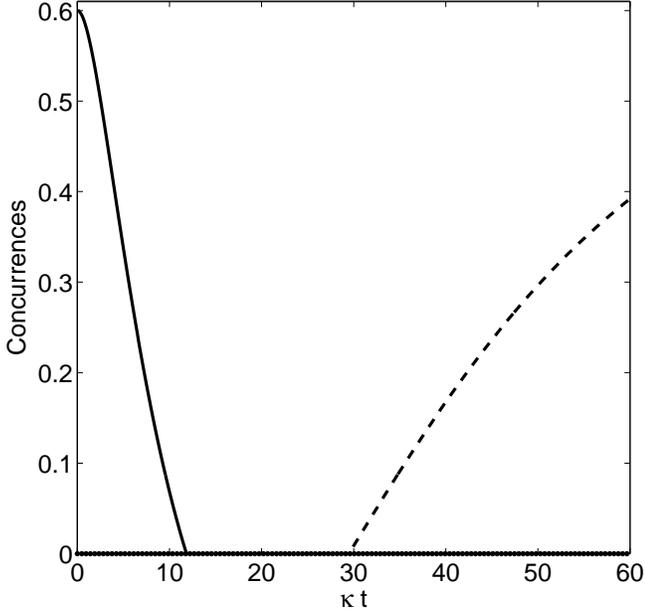}
\caption{Evolution of concurrence of the subsystems $a_1\otimes a_2$ (solid line),  $c_1\otimes c_2$ (circles), $r_1\otimes r_2$ (dashed line) for $\beta=3 \alpha$ and $\gamma=0.1$.}
\label{tw}
\end{figure}

In this time window, where no entanglement is found in subsystems $a_1 \otimes a_2$, $c_1 \otimes c_2$ and $r_1 \otimes r_2$, the question of where the entanglement goes in this time window becomes relevant. In order to answer this question we must analyze the entanglement present between other
bipartite subsystems of the overall system. 

First we consider the subsystem $(a_1,c_1,r_1)\otimes
(a_2,c_2,r_2)$. Since this bipartite subsystem is in a pure state at all times, we are able to quantify the entanglement  through the square root of the
\emph{tangle}~\cite{Rungta03}. This entanglement is given by:
\begin{equation}
\sqrt{\tau _{12}(t)}=2\alpha\beta, \label{tangle12}
\end{equation}
which corresponds to the same amount of entanglement initially
present in the atomic subsystem. This conservation of the global
entanglement occurs due to the non-interacting character of the subsystem $(a_1,c_1,r_1)\otimes
(a_2,c_2,r_2)$
\begin{figure}[t]
\includegraphics[width=80mm]{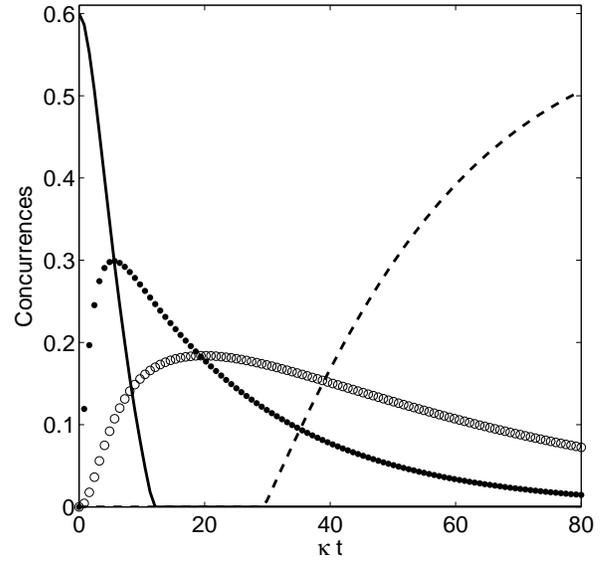}
\caption{Evolution of concurrence of the subsystems $a_1\otimes c_1$(dots),  $c_1\otimes r_1$ (circles), $a_1\otimes a_2$ (solid line) and $r_1\otimes r_2$ (dashed line) for $\beta=3 \alpha$ and $\gamma=0.1$.}
\label{Figa1c1c1r1}
\end{figure}
\begin{figure}[t]
\includegraphics[width=80mm]{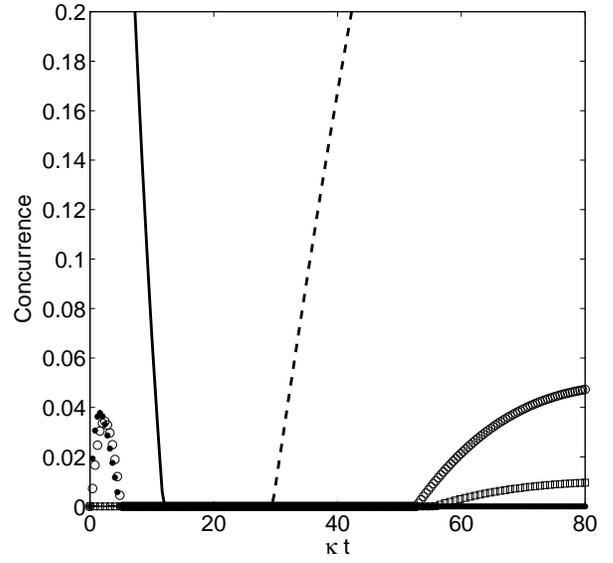}
\caption{Evolution of concurrence of the subsystems $a_1\otimes c_2$ (dots),  $a_1\otimes r_2$ (circles), $c_1\otimes r_2$ (squares), $a_1\otimes a_2$ (solid line) and $r_1\otimes r_2$ (dashed line) for $\beta=3 \alpha$ and $\gamma=0.1$.}
\label{noninteracting}
\end{figure}
For interacting subsystems, for example, $a_{1(2)}\otimes c_{1(2)}$
and $c_{1(2)}\otimes r_{1(2)}$, concurrences are shown in
Fig.~\ref{Figa1c1c1r1}. From the figure, we observe that in the time window where no entanglement is found in subsystems $a_1 \otimes a_2$ and $r_1 \otimes r_2$, the interacting subsystems show entanglement. Indeed, in such a time window, the entanglement seems to flow only through interacting subsystems. To confirm this, we need to consider different non-interacting subsystems such as, $a_1\otimes c_2$, $a_1\otimes r_2$ and $c_1\otimes r_2$. In these cases the concurrences are respectively given by:
\begin{eqnarray}
\mathcal{C}_{a_{1}c_{2}}(t) &=& \max\{0,2\big[\alpha\beta E_t |G_t |-\sqrt{w(E_t, |G_t |)}\big]\}, \\
\mathcal{C}_{a_{1}r_{2}}(t) &=& \max\{0,2 \big[\alpha\beta E_t R_t -\sqrt{w(E_t, R_t)}\big]\}, \\
\mathcal{C}_{c_{1}r_{2}}(t) &=& \max\{0,2 \big[\alpha\beta |G_t |
R_t-\sqrt{w(|G_t |, R_t)}\big]\},
\end{eqnarray}
where $w(x,y)=\alpha^4 xy^2 (1-x^2)(1-y^2)$. The temporal evolution of these concurrences is shown in Fig.~\ref{noninteracting}. According to this figure, in the aforementioned time window, no contribution to entanglement comes from such non-interacting subsystems. Moreover, the time window where no entanglement is found in this subsystems is longer than the respective time window for the non-interacting subsystems $a_1 \otimes a_2$ and $r_1 \otimes r_2$.
This confirms that in this finite time window the entanglement can flow only through interacting subsystems, such as $a_{1(2)}\otimes c_{1(2)}$ and $c_{1(2)}\otimes r_{1(2)}$, while the non-interacting subsystems have no entanglement.

\section{summary}
In summary, we have studied the dynamics of the entanglement transfer of two uncoupled systems each composed of a single atom inside a leaky cavity.  We have developed a hybrid analytical approach to find the entanglement dynamics without tracing out reservoir modes, allowing us to study the entanglement behavior in different subsystems. In particular, we have found that the entanglement initially located in two atoms is asymptotically mapped to the reservoirs degrees of freedom. Moreover, although reservoirs are connected to atoms through the cavities, we show that for a set of initial amplitudes and coupling constants, there are entangled and unentangled phases in the cavities. We have also shown that there is a time window where no entanglement is found in non-interacting subsystems such us atoms, cavities or reservoir. In this case, we observe that the entanglement flows only through interacting subsystems.

CEL acknowledge financial support from Fondecyt 11070244 and PBCT-CONICYT PSD54, GR from Juan de la Cierva Program, JCR from Fondecyt 1070157. CEL and JCR akcnowledge support from Financiamiento Basal para Centros Cient\'iÞcos y Tecnol\'ogicos de Excelencia.

\end{document}